\documentclass{iaus}
\usepackage{graphicx}

\newcommand{\Bf}{{\bf B}}

\newcommand{\vf}{{\bf v}}
\newcommand{\pa}{\partial}

\title[Solar cycle memory] %% give here short title %%
{Forecasting the solar activity cycle: new
insights}
\author[D. Nandy \& B.\ B. Karak]   %% give here short author list %%
{Dibyendu Nandy$^1$
 \and Bidya Binay Karak$^2$}

\affiliation{$^1$Indian Institute for Science Education and Research, Kolkata, Mohampur 741252, West~Bengal, India\\Email: {\tt dnandi@iiserkol.ac.in}\\
$^2$Department of Physics, Indian Institute of Science, Bangalore 560012, India}
\pubyear{2013}
\volume{294}  %% insert here IAU Symposium No.
\pagerange{439--444}
\jname{Solar and Astrophysical Dynamos and Magnetic Activity}
\editors{A.G. Kosovichev, E.M. de Gouveia Dal Pino, \& Y.Yan, eds.}
\begin{document}
\maketitle
\begin{abstract}
Having advanced knowledge of solar activity is important because the Sun's magnetic
output governs space weather and impacts technologies reliant on space. However, the irregular
nature of the solar cycle makes solar activity predictions a challenging task. This is best achieved
through appropriately constrained solar dynamo simulations and as such the first step towards
predictions is to understand the underlying physics of the solar dynamo mechanism. In Babcock--Leighton type dynamo models, the poloidal field is generated near the solar surface whereas the
toroidal field is generated in the solar interior. Therefore a finite time is necessary for the coupling
of the spatially segregated source layers of the dynamo. This time delay introduces a memory
in the dynamo mechanism which allows forecasting of future solar activity. Here we discuss how
this forecasting ability of the solar cycle is affected by downward turbulent pumping of magnetic
flux. With significant turbulent pumping the memory of the dynamo is severely degraded and
thus long term prediction of the solar cycle is not possible; only a short term prediction of the
next cycle peak may be possible based on observational data assimilation at the previous cycle
minimum.
\end{abstract}
\firstsection % if your document starts with a section,
\section{Introduction}
The solar cycle is not regular. The individual cycles vary in strength from one cycle to
another. Therefore prediction of future cycles is a non-trivial task. However forecasting
future cycle amplitudes is important because of the impact of solar activity on our space
environment. Unfortunately, recent efforts to predict the solar cycle did not reach any
consensus, with a wide range of forecasts for the strength of the ongoing cycle 24 (Pesnell
2008).

Kinematic dynamo models based on the Babcock-Leighton mechanism has proven to be
a viable approach for modeling the solar cycle
(e.g., Mu\~noz-Jaramillo et al. 2010; Nandy 2011; Choudhuri 2013). In such models, the poloidal field is generated from the decay
of tilted active regions near the solar surface mediated via near-surface flux transport
processes. In this model the large-scale coherent meridional circulation plays a crucial
role (Choudhuri et al. 1995; Yeates, Nandy \& Mackay 2008; Karak 2010; Nandy, Mu\~noz-Jaramillo \& Martens 2011; Karak \& Choudhuri 2012). 
This is because the meridional
circulation is believed to transport the poloidal field -- generated near the solar surface
-- to the interior of the convection zone where the toroidal field is generated through
stretching by differential rotation. The time necessary for this transport introduces a
memory in the solar dynamo, i.e.,\ the toroidal field (which gives rise the sunspot eruptions) has an in-built ``memory'' of the earlier poloidal field. Yeates, Nandy \& Mackay (2008) systematically studied this 
issue and showed that in the advection-dominated
regime of the dynamo the poloidal field is mainly transported by the meridional circulation and the solar cycle memory persists over many cycles (see also Jiang, Chatterjee \& Choudhuri 2007). On the other hand, in the diffusion-dominated regime of the dynamo,
the poloidal field is mainly transported by turbulent diffusion and the memory of the
solar cycle is short -- roughly over a cycle. Recent studies favor the diffusion-dominated
solar convection zone (Miesch et al.\ 2011) and the diffusion-dominated dynamo is successful in modeling many important aspects of the solar cycle including the
the Waldmeier effect and the grand minima (Karak \& Choudhuri 2011; Choudhuri \& Karak 2009; Karak
2010; Choudhuri \& Karak 2012; Karak \& Petrovay 2013).
Using an advection dominated B-L dynamo Dikpati de Toma \& Gilman (2006) predicted a strong cycle 24. On the other hand, Choudhuri et al. (2007) used a diffusion-dominated model and predicted a weak cycle (see also Jiang et al.\ 2008). However in most
of the models, particularly in these prediction models, the turbulent pumping of magnetic 
flux -- an important mechanism for transporting magnetic field in the convection
zone -- was ignored. Theoretical as well as numerical studies have shown that a horizontal
magnetic field in the strongly stratified turbulent convection zone is pumped preferentially 
downward towards the base of the convection zone (stable layer) and a few m/s
pumping speed is unavoidable in many convective simulations (e.g., Petrovay \& Szakaly
1993; Brandenburg et al.\ 1996; Tobias et al.\ 2001; Dorch \& Nordlund 2001; Ossendrijver
et al. 2002; K\"apyl\"a et al.\ 2006; Racine et al.\ 2011). Recently, we have studied the impact
of turbulent pumping on the memory of the solar cycle and hence its relevance for solar
cycle forecasting (Karak \& Nandy 2012). Here we provide a synopsis of our findings and
discuss its implications for solar cycle predictability.

\section{Model}
The evolution of the magnetic fields for a kinematic $\alpha-\Omega$ dynamo model is governed by
the following two equations.
\begin{equation}
\frac{\pa A}{\pa t} + \frac{1}{s}(\vf.\nabla)(s A)
= \eta_{p} \left( \nabla^2 - \frac{1}{s^2} \right) A + \alpha B~~~~~~~~~~~~~~~~~~~~~~~~~~~~~~~~~~~~~~~~~~~~~~~~~
\label{pol_eq}
\end{equation}
\begin{equation}
\frac{\pa B}{\pa t}
+ \frac{1}{r} \left[ \frac{\pa}{\pa r}
(r v_r B) + \frac{\pa}{\pa \theta}(v_{\theta} B) \right]
= \eta_{t} \left( \nabla^2 - \frac{1}{s^2} \right) B + s(\Bf_p.\nabla)\Omega + \frac{1}{r}\frac{d\eta_t}{dr}
\frac{\partial}{\partial{r}}(r B)~~
\end{equation}
with $s = r \sin \theta$. Here $A$ is the vector potential of the 
poloidal magnetic field, $B$ is the toroidal magnetic field, $\vf=v_r \hat{r} + 
v_{\theta} \hat{\theta}$ is the meridional 
circulation, $\Omega$ is the internal angular velocity, $\alpha$ is the source term for the poloidal field by the B-L mechanism and $\eta_p$, $\eta_t$
are the turbulent diffusivities for the poloidal and toroidal 
components. With the given ingredients, we solve the above two equations
to study the evolution of the magnetic field in the dynamo model.
The details of this model can be found in Nandy \& Choudhuri (2002) and Chatterjee, Nandy \& Choudhuri (2004).
However for the sake of comparison with the earlier results we use the exactly same parameters as given in Yeates, Nandy \& Mackay (2008).

In the mean-field induction equation, the turbulent pumping naturally appears as an advective term.
Therefore to include its effect in the present dynamo model, we include the turbulent pumping term shown by
the following expression in the advection term of the poloidal field equation (Eq.~\ref{pol_eq}).\\
\begin{equation}
\gamma_r = - \gamma_{0r} \left[ 1 + \rm{erf}\left( \frac{r - 0.715}{0.015}\right) \right] \left[ 1 - \rm{erf} \left( \frac{r-0.97}{0.1}\right) \right] 
 \left[ \rm{exp}\left( \frac{r-0.715}{0.25}\right) ^2 \rm{cos}\theta +1\right],\\
%\label{rpumping}
\end{equation}
where $\gamma_{0r}$ determines the strength of the pumping what 
we vary in our simulations. 
Note that we introduce pumping only in the poloidal field because turbulent pumping is 
likely to be relatively less effective on the toroidal component (e.g., K\"apyl\"a et al.\ 2006). The
toroidal field is stronger, intermittent and subject to buoyancy forces and therefore it is
less prone to be pumped downwards. Also note that we do not consider any latitudinal
pumping.

To study the solar cycle memory we have to make the strength of the cycle 
unequal by introducing some stochasticity in the model. Presently we believe
that there are two important sources of randomness in the flux transport dynamo model -- the 
stochastic fluctuations in the B-L process of generating the poloidal field and the 
stochastic fluctuations in the meridional circulation.
In this work, we introduce stochastic fluctuations in the $\alpha$ appearing in Eq.~\ref{pol_eq} to capture the
irregularity in the B-L process of poloidal field generation. We set  
$\alpha_0 = \alpha_{\rm{base}} + \alpha_{\rm{fluc}} \sigma (t, \tau_{\rm{cor}})$. 
Throughout all the calculations we take $\alpha_{\rm{fluc}} = \alpha_{\rm{base}} = 30$~m~s$^{-1}$ 
(i.e., $100\%$ level of fluctuations).
The coherence time $\tau_{cor}$ is chosen in such a way 
that there are around 10 fluctuations in each cycle. 
%In doing so, the value of $\tau_{cor}$ always lies within 0.5 yr to 2.0 yr which is in consistent with the surface flux transport process in B-L effect.

\section{Results}
We have carried out extensive simulations with stochastically varying $\alpha$ at 
different downward pumping speed varied from 0 to 4 m s$^{-1}$. We have performed simulations in two different 
regimes of dynamo---the diffusion-dominated regime with parameters $v_0 = 
15$~m~s$^{-1}$, $\eta_0 = 1 \times 10^{12}$~cm$^2$~s$^{-1}$ and the advection-dominated 
regime with $v_0 = 26$~m~s$^{-1}$, $\eta_0 = 1\times 10^{12}$~cm$^2$~s$^{-1}$. 
In the previous case the diffusion of the fields are more important compared to the
advection by meridional flow whereas in the latter case it is the other way round.

\begin{figure}
% \vspace*{-2.0 cm}
\begin{center}
 \includegraphics[width=4.0in]{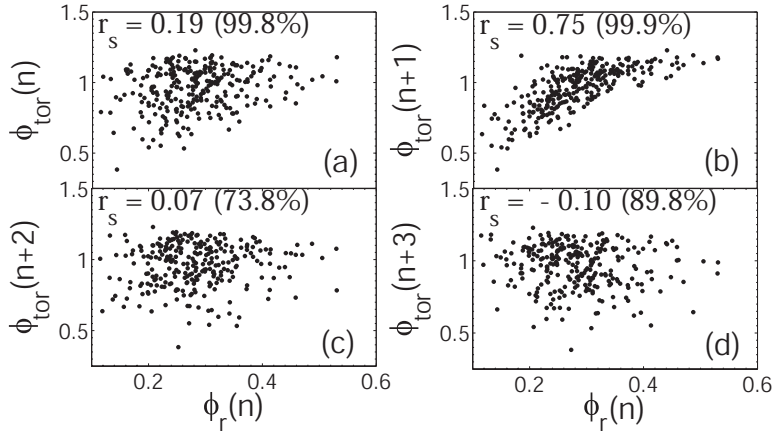}
\caption{Scattered plots of the peak (polar) radial flux $\phi_{\rm{r}}(n)$ and the peak (deep-seated) toroidal flux $\phi_{\rm{tor}}$ of cycle (a) $n$ (b) $n+1$, (c) $n+2$, and (d) $n+3$ in the advection-dominated regime with a pumping speed amplitude of 2 m~s$^{-1}$. The flux values are in units of $10^{25}$ Mx. The Spearman's rank correlation coefficients ($r_s$) along with significance levels are inscribed. Reproduced from Karak \& Nandy (2012).}
   \label{corr_ap2}
\end{center}
\end{figure}

\begin{figure}
% \vspace*{-2.0 cm}
\begin{center}
 \includegraphics[width=4.0in]{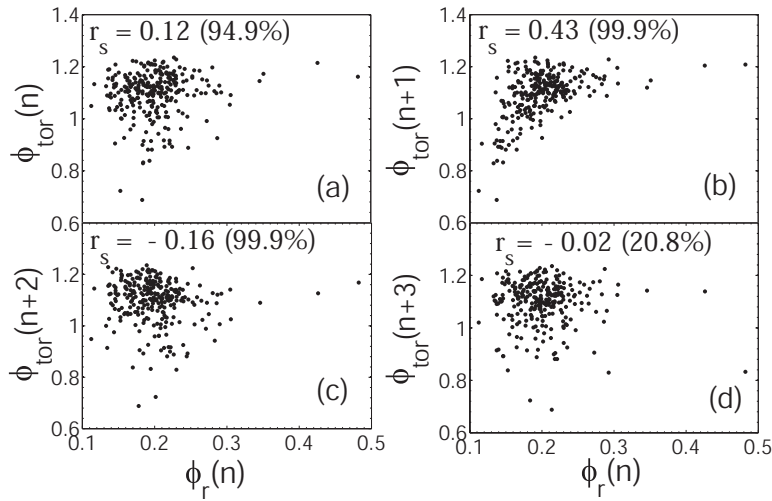}
% \vspace*{-1.0 cm}
 \caption{Same as Fig.~1, but for the diffusion-dominated regime. Reproduced from Karak \& Nandy (2012).}
   \label{corr_dp2}
\end{center}
\end{figure}

Other than some obvious effects of the turbulent pumping on the solar cycle period and 
the latitudinal distribution of the magnetic field (which have already been explored by Guerrero \& de Gouveia Dal Pino 2008) 
we are interested here to see the 
dependence of the toroidal field on the previous cycle poloidal fields. To do this
we compute the correlation between the peak of the surface radial flux ($\phi_{r}$) of cycle $n$ 
with that of the deep-seated toroidal flux ($\phi_{tor}$) of different cycles. Here we consider $\phi_{r}$ as 
the flux of radial field over the 
solar surface from latitude $70^0$ to $89^0$, and $\phi_{tor}$ as the flux of 
toroidal field over the region $r=0.677-0.726 R$ and latitude $10^0$ to $45^0$. 
In table~1, we present the Spearman's rank correlation coefficients 
and significance levels in two different regimes with increasing pumping speed. 
From this table we see that in the advection-dominated 
regime, in absence of pumping, the polar flux of cycle $n$ correlates with 
the toroidal flux of cycle $n+1$, $n+2$ and $n+3$, whereas in diffusion dominated regime, 
only one cycle correlation exist (i.e., the polar flux of cycle $n$ only correlates 
with the toroidal flux of cycle $n+1$). This is consistent with 
Yeates, Nandy \& Mackay (2008). However, it is interesting to see that 
with the increase of the pumping speed in the advection-dominated region, the higher order 
correlations slowly diminish and even just at 2.0~m~s$^{-1}$ pumping speed only the $n$ to $n+1$ correlation exists and other correlations
have destroyed. However the behavior in the diffusion-dominated regime remains qualitatively unchanged. 
Fig.~\ref{corr_ap2} shows the correlation plot with 2.0~m~s$^{-1}$ pumping amplitude for 
the advection-dominated regime whereas Fig.~\ref{corr_dp2} shows the same for the diffusion-dominated case.
 
Another important result of these analyses is that with the increase of the 
strength of the pumping the $n$ to $n+1$ correlations are also decreasing rapidly 
in both the advection-dominated and in the diffusion-dominated regime (see Table~1).

\begin{table}[t]
\caption[]{Correlation coefficients ($r_s$) and percentage significance levels ($p$) for peak surface radial flux $\Phi_{\rm{r}}$ of cycle $n$ versus peak toroidal flux $\Phi_{\rm{tor}}$ of different cycles for 275 solar cycles data. The first column denotes the amplitude of the turbulent pumping speed in various simulation studies. The top row corresponds to the case without turbulent pumping and subsequent rows correspond to simulations with increasing pumping speeds.}
  \begin{center}\begin{tabular}{lrrrrr}
 \hline
                               &              &Dif. Dom.                                       & Adv. Dom.      \\
 \hline
Pumping                        & Parameters~~~~~~~                                 & $r_s$ ($p$)& $r_s$~($p$) \\
\hline
                               & $\Phi_{\rm{r}}(n)~\&~\Phi_{\rm{tor}}(n)$~~~~~~& $0.19~(99.9)$ & $0.57~(99.9)$\\
                               & $\Phi_{\rm{r}}(n)~\&~\Phi_{\rm{tor}}(n+1)$    & $0.64~(99.9)$ & $0.77~(99.9)$\\ [-1ex]
\raisebox{1.0ex}{0 m s$^{-1}$} & $\Phi_{\rm{r}}(n)~\&~\Phi_{\rm{tor}}(n+2)$    & $0.04~(55.9)$ & $0.46~(99.9)$\\
                               & $\Phi_{\rm{r}}(n)~\&~\Phi_{\rm{tor}}(n+3)$    & $0.22~(99.9)$ & $0.27~(99.9)$\\
\hline
                               & $\Phi_{\rm{r}}(n)~\&~\Phi_{\rm{tor}}(n)$~~~~~~& $-0.06~(67.0)$ & $0.41~(99.9)$\\
                               & $\Phi_{\rm{r}}(n)~\&~\Phi_{\rm{tor}}(n+1)$    & $0.67~(99.9)$ & $0.72~(99.9)$\\[-1ex]
\raisebox{1.0ex}{1 m s$^{-1}$} & $\Phi_{\rm{r}}(n)~\&~\Phi_{\rm{tor}}(n+2)$    & $0.09~(83.9)$ & $0.29~(99.9)$\\
                               & $\Phi_{\rm{r}}(n)~\&~\Phi_{\rm{tor}}(n+3)$    & $-0.02~(26.5)$ & $-0.01~(18.9)$\\
\hline
                               & $\Phi_{\rm{r}}(n)~\&~\Phi_{\rm{tor}}(n)$~~~~~~& $0.12~(94.9)$ & $0.19~(99.8)$\\
                               & $\Phi_{\rm{r}}(n)~\&~\Phi_{\rm{tor}}(n+1)$    & $0.43~(99.9)$ & $0.75~(99.9)$\\[-1ex]
\raisebox{1.0ex}{2 m s$^{-1}$} & $\Phi_{\rm{r}}(n)~\&~\Phi_{\rm{tor}}(n+2)$    & $-0.16~(99.9)$ & $0.07~(73.8)$\\
                               & $\Phi_{\rm{r}}(n)~\&~\Phi_{\rm{tor}}(n+3)$    & $-0.02~(20.8)$ & $-0.10~(89.8)$\\
\hline
                               & $\Phi_{\rm{r}}(n)~\&~\Phi_{\rm{tor}}(n)$~~~~~~& $0.11~(49.2)$ & $0.29~(92.0)$\\
                               & $\Phi_{\rm{r}}(n)~\&~\Phi_{\rm{tor}}(n+1)$    & $0.32~(99.9)$ & $0.62~(99.9)$\\[-1ex]
\raisebox{1.0ex}{3 m s$^{-1}$} & $\Phi_{\rm{r}}(n)~\&~\Phi_{\rm{tor}}(n+2)$    & $-0.18~(99.6)$ & $0.07~(78.0)$\\
                               & $\Phi_{\rm{r}}(n)~\&~\Phi_{\rm{tor}}(n+3)$    & $0.03~(36.6)$ & $-0.10~(91.6)$\\
\hline
                               & $\Phi_{\rm{r}}(n)~\&~\Phi_{\rm{tor}}(n)$~~~~~~& $0.19~(99.8)$ & $0.30~(99.9)$\\
                               & $\Phi_{\rm{r}}(n)~\&~\Phi_{\rm{tor}}(n+1)$    & $0.26~(99.9)$ & $0.46~(99.9)$\\[-1ex]
\raisebox{1.0ex}{4 m s$^{-1}$} & $\Phi_{\rm{r}}(n)~\&~\Phi_{\rm{tor}}(n+2)$    & $-0.16~(99.3)$ & $0.07~(72.8)$\\
                               & $\Phi_{\rm{r}}(n)~\&~\Phi_{\rm{tor}}(n+3)$    & $-0.10~(91.9)$ & $-0.22~(99.9)$\\
\hline
\end{tabular}
  \end{center}
\end{table}

\section{Conclusion and Discussion}
We have introduced turbulent pumping of the magnetic flux in a B-L type kinematic
dynamo model and have carried out several extensive simulations with stochastic fluctuation in the B-L $\alpha$ with different strengths of downward turbulent pumping in both
advection- and diffusion-dominated regimes of the solar dynamo. We find that multiple
cycle correlations between the surface polar flux and the deep-seated toroidal flux in
the advection-dominated dynamo model degreades severely when we introduce turbulent
pumping. With 2~m~s$^{-1}$ as the typical pumping speed, the timescale for the poloidal
field to reach the base of the convection zone is about 4 years, which is even shorter than
the timescale of turbulent diffusion (and much shorter than the advective timescale due
to meridional circulation). Consequently the behavior found in the advection-dominated
dynamo model with pumping is similar to that seen in the diffusion-dominated dynamo
model indicating that downward turbulent pumping short-circuits the meridional flow
transport loop for the poloidal flux. This transport loop is first towards the poles at
near-surface layers and then downwards towards the deeper convection zone and subsequently 
equatorwards. However, when pumping is dominant, then the transport loop is
predominantly downwards straight into the interior of the convection zone.

An interesting and somewhat counter-intuitive possibility that our findings raise is
that the solar convection zone may not be diffusion-dominated, or advection-dominated,
but rather be dominated by turbulent pumping. Note that this does not rule out the
possibility that in the stable layer beneath the base of the convection zone, meridional
circulation still plays an important and dominant role in the equatorward transport of
toroidal flux and thus, in generating the butterfly diagram.

Our result implies with turbulent pumping as the dominant mechanism for flux transport, the solar cycle memory is short. This short memory, lasting less than a complete 11
year cycle implies that solar cycle predictions for the maxima of cycles are best achieved
at the preceding solar minimum, about 4-5 years in advance and long-term predictions
are unlikely to be accurate. This also explains why early predictions for the amplitude
of solar cycle 24 were inaccurate and generated a wide range of results with no consensus. 
The lesson that we take from this study is that it is worthwhile to invest time and
research to understand the basic physics of the solar cycle first, and that advances made
in this understanding will lead to better forecasting capabilities for solar activity.

\begin{acknowledgements}
The authors would like to thank the Ministry of Human Resource Development and the
Department of Science and Technology, Government of India, for funding this research
summarized here.

\end{acknowledgements}

\end{document}